\documentstyle[12pt,epsf]{article}
\let\section=\subsection     \let\subsection=\subsubsection                

\begin{document}
\begin{center}
   {\large \bf MANY-BODY THEORY OF NUCLEAR AND}\\[2mm]
   {\large \bf NEUTRON STAR MATTER}\\[5mm]
   V. R. PANDHARIPANDE, A. AKMAL and D. G. RAVENHALL \\[5mm]
 {\small \it  Department of Physics, University of Illinois at Urbana-Champaign \\
1110 W. Green St., Urbana, Illinois 61801, USA \\[8mm] }
\end{center}

\begin{abstract}\noindent
We present results obtained for nuclei, nuclear and neutron star matter, 
and neutron star structure obtained with the recent Argonne $v_{18}$ two-
nucleon and Urbana IX three-nucleon interactions including relativistic boost 
corrections.  These interactions predict that matter will undergo a 
transition to a spin layered phase with neutral pion condensation. We 
also consider the possibility of a transition to quark matter.
\end{abstract}

\section{Introduction}
Properties of matter having densities up to $\sim 1$ nucleon$/$fm$^3$ 
are important in determining the structure and maximum mass of neutron 
stars.  At such densities the average interparticle spacing is $\sim 1$ 
fm. It is well known that the rms charge radius of proton is $\sim 0.8$
fm, thus one may be concerned that nucleons 1 fm apart have too
large a structural overlap, and may even cease to be nucleons. 
However, the large charge radius of protons is due to a  
diffuse cloud attributed to mesons.  The charge form factor of the proton, as 
well as the magnetic form factors of the proton and the 
neutron are well approximated 
by the dipole form:
\begin{equation}
F(q) = (1+q^2/q^2_0)^{-2};\ \ \ q_0=0.84\ \frac{GeV}{c}.
\label{eq:ffch}
\end{equation}
The charge density obtained by inverting this form factor is:
\begin{equation}
\rho_{ch}(r) = 3.3\ e^{-r/0.23}\ {\rm fm}^{-3},\ \ r\ {\rm in\ fm}.
\label{eq:rhoch}
\end{equation}
The charge densities of two protons placed at $\pm 0.5$ fm on the z-axis 
are shown in fig.1.  We see that there is not too much structural overlap 
of protons 1 fm apart.  We may therefore assume that they still are 
protons, 
and absorb the effects of the overlap into the two nucleon interaction. 
This assumption is supported by studies of the deuteron.
The deuteron wave function peaks at $\sim 1$ fm separation between the 
neutron and the proton in agreement with the observed deuteron form 
factors \cite{FPPWSA 96}.

\begin{figure}[htpb]
\centerline{\mbox{\epsfysize=7cm \epsffile{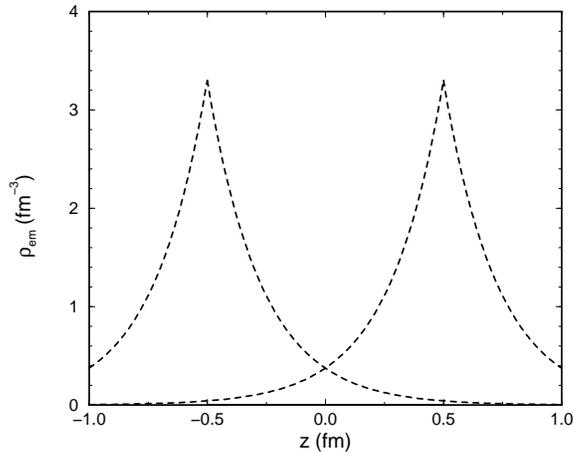}}}
\caption{Charge density of two nucleons at $z=\pm 0.5$ fm, along the z-axis}
\end{figure}

In nuclear many-body theory, nuclei and nuclear matter are described by the 
Hamiltonian:
\begin{equation}
H = \sum_i \frac{-\nabla^2}{2m} + \sum_{i<j}(v_{ij}+\delta v({\bf P}_{ij}))
+\sum_{i<j<k} V_{ijk},
\label{eq:ham}
\end{equation}
in which $i,j,k,...$ denote nucleons.  The first term is the nonrelativistic 
kinetic energy, and the second contains the two nucleon interaction $v_{ij}$ in 
the center of mass frame of the interacting pair.
The relativistic boost interaction 
$\delta v({\bf P}_{ij})$ describes the dependence of two-nucleon interaction 
on their total center of mass momentum ${\bf P}_{ij}$, and the last term 
contains the three nucleon interaction.  In principle the $H$ can have 
additional terms, such as the 
boost correction for three-nucleon interaction,  
four nucleon interaction, {\em etc.}; however these terms 
are believed to have 
a negligible effect on neutron star structure.  
The above $H$ describes only the hadronic part of neutron star matter which
also contains electrons and, at high densities, muons to preserve 
charge neutrality and beta equilibrium.  The relativistic kinetic energies 
of the leptons and the Coulomb interaction energies are to be added to the
above hadronic $H$.

Recent developments in various terms of the nuclear Hamiltonian are 
discussed in the next section, while progress in the variational 
calculations of the properties of matter from this $H$ is reviewed in 
sect. 3.  The results for neutron stars are given in sect. 4, while 
the possibility of a phase transition to quark matter is considered in 
sect. 5. A more detailed description of 
this work will be published elsewhere.

\section{Modern Models of Nuclear Forces}

In the early 1990's the Nijmegen group \cite{SKRD 93} 
carefully examined all the data on NN 
scattering at energies below 350 MeV published between 1955 and 1992.  They 
extracted 1787 proton-proton and 2514 proton-neutron ``reliable'' data, and 
showed that these could determine all NN scattering phase shifts and 
mixing parameters quite accurately.  NN interaction models which fit this 
Nijmegen data base with a ${\chi}^2/N_{data} \sim 1$ are termed ``modern''. 
These include the Nijmegen models \cite{SKTR 94} called Nijmegen I, II and 
Reid 93, the Argonne $v_{18}$ \cite{WSS 95} (A18) and CD-Bonn \cite{MSS 96}. 
In order to fit both the proton-proton and neutron-proton scattering data 
simultaneously and accurately, these models include a detailed description 
of the electromagnetic interactions and terms that violate the isospin 
symmetry of the strong interaction via the differences in the masses of the 
charged and neutral pions, etc.

\begin{figure}[htpb]
\centerline{\mbox{\epsfysize=7cm \epsffile{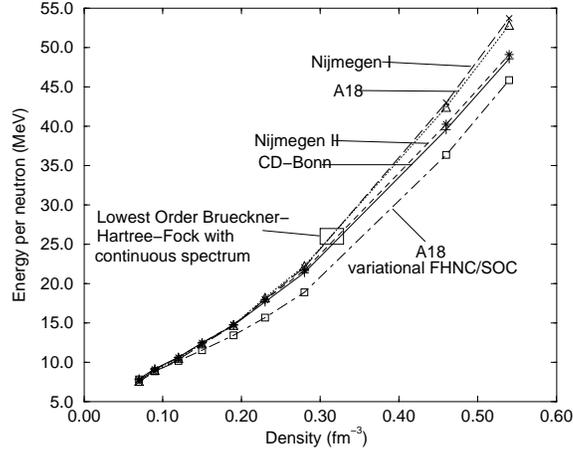}}}
\vspace{-0.5cm}
\caption{The $E(\rho)$ of neutron matter calculated from modern potential models.  The upper four curves show results obtained with Brueckner-Hartree-Fock calculations, while the lower curve is with variational calculations.}
\end{figure}

In these five models the two nucleon interaction $v_{ij}$ is expressed as 
a sum of the one-pion exchange potential $v^{\pi}_{ij}$ and the rest of 
the interaction $v^R_{ij}$.  They use
different parameterizations of the $v^R_{ij}$, and the 
Nijmegen-I and CD-Bonn also include nonlocalities suggested by 
boson-exchange representations.  Thus, like the older models, they make 
different predictions for many-body systems.  However, the differences in 
their predictions are much smaller than those between older models, presumably 
because they exactly fit the same large data set.  For example, the triton 
ground state energies predicted by the modern Nijmegen and Argonne models,
without the $\delta v({\bf P}_{ij})$ and $V_{ijk}$ terms in the Hamiltonian, 
are between -7.62 and -7.72 MeV \cite{FPSS 93}, while that of CD-Bonn is -8.00 
MeV \cite{MSS 96}.  This should be compared with the spread from -7.35 MeV
(original Reid) to -8.35 MeV (Bonn-A) in the predictions of the older models.
The predictions for the energies of dense neutron matter obtained from the 
modern potentials using lowest order Brueckner-Hartree-Fock (LOB) method 
\cite{EHMMP 97} are also quite close together as shown in fig. 2.  
This figure also shows the energies of neutron matter calculated from the 
A18 potential by the variational method using chain summation 
techniques \cite{AP 97}.  The difference between the results obtained for 
this modern potential using LOB and variational methods is 
larger than that in the results of all the modern potentials with the 
LOB method. 

The modern potentials, like their predecessors,
do not explain all of the observed nuclear properties. 
They all underbind the triton, and 
overpredict the density of nuclear matter.  
It is necessary to add three-nucleon 
(NNN) interactions to the Hamiltonian to reproduce these observables.  The 
Urbana models of NNN interaction have only two terms:
\begin{equation}
V_{ijk} = V^{2{\pi}}_{ijk} + V^R_{ijk} ,
\label{eq:vijk}
\end{equation}
the first gives the two-pion exchange NNN interaction via the pion-nucleon 
delta resonance \cite{FM 57}, and 
its strength is denoted by the parameter $A_{2{\pi}}$; the second is
a phenomenological spin-isospin independent, shorter range term with strength 
denoted by $U_0$.  
The values of the parameters 
$A_{2{\pi}}$ and $U_0$ of the Urbana-IX (UIX) NNN potential are chosen such 
that in combination with A18 NN potential triton energy is
reproduced via exact calculations and the equilibrium density of nuclear 
matter is reproduced via variational calculations \cite{PPCPW 97}.  
The boost interaction $\delta v({\bf P}_{ij})$ was negelected in these 
calculations to determine $A_{2\pi}$ and $U_0$.  Its effect is discussed 
later. It is unlikely that a combination of CD-Bonn and UIX potentials 
will reproduce the triton energy.  Parameters of the NNN interaction 
have to refitted for each model of the NN interaction, as in the recent 
work of the Bochum group \cite{WG 97} using the Tucson-Melbourne \cite{TMPOT}
model of $V^{2{\pi}}_{ijk}$.  The fitted $V_{ijk}$ then 
partly corrects for the deviation of the model $v_{ij}$ from its
exact representation.  We hope that predictions of the modern combinations 
of $v_{ij}$ and $V_{ijk}$ will be significantly less model dependent than those 
of the $v_{ij}$ alone.

It has recently become possible to calculate all the bound states of up to 
eight nucleons from realistic nuclear forces with the Greens Function 
Monte Carlo (GFMC)
method. Since solar and primodial fusion reactions primarily involve nuclei
having $A \leq 8$, they have a special role in the universe.  Here we use
the results obtained with A18, without $\delta v({\bf P}_{ij})$, 
and UIX interactions to test the accuracy 
of variational wave functions and the interaction models.  The calculated 
energies from \cite{PPCPW 97} and \cite{W 98} are listed in Table I.
Its first two columns specify the nuclear state and 
the next four give the calculated expectation values of the interaction 
components.  The last three columns list the energy calculated with GFMC,
the difference between the experimental and GFMC energies, and that 
between our optimum variational 
and GFMC energies.  The $v^{\pi}_{ij}$ seems to give 
the dominant contribution to nuclear binding, while $v^R_{ij}$ is also 
essential for all but the deuteron.  There is a large cancellation 
between the kinetic and two-nucleon interaction contributions, 
causing the total nuclear 
energy to be much smaller than the $\langle v_{ij} \rangle$.  The 
$\langle V^{2{\pi}}_{ijk} \rangle$ is much smaller than $\langle v^{\pi}_{ij} 
\rangle$; however, it is a significant fraction of the nuclear binding energy. 
The $\langle V^R_{ijk} \rangle$ is the smallest, and $\Delta E_{expt.}$, the 
difference between experiment and theory is even smaller.  In the neutron 
rich nuclei $^7$He and $^8$He, the $\Delta E_{expt.}$ is comparable to the 
$\langle V^R_{ijk} \rangle$, indicating that the UIX model may not be
describing the interaction between three neutrons very well.  All the p-shell 
nuclei having $A > 5$ are underbound, and the problem increases with the 
magnitude of nuclear isospin 
T.  New models of $V_{ijk}$ are being studied to reduce the $\Delta E_{expt.}$.

\begin{table}
\caption{Results of Quantum Monte Carlo Calculations in MeV}
\vspace{0.2cm}
\begin{tabular}{rlrrrrrrr}
\hline
$^A$Z & $(J^{\pi};T)$   &  $v^{\pi}_{ij}$  &  $V^{2{\pi}}_{ijk}$  &  $v^R_{ij}$  &  $V^R_{ijk}$  &  $E_{GFMC}$  &  ${\Delta}E_{expt.}$  &  ${\Delta}E_{VMC}$\\
\hline
$^2$H & $(1^+;0)$ & -21.3 & 0 & -0.8 & 0 & -2.22 & 0 & 0 \\
$^3$H & $({\frac{1}{2}}^+;\frac{1}{2})$ & -43.8 & -2.2 & -14.6 & 1.0 & -8.47 & -0.01(1) & 0.15 \\
$^4$He & $(0^+;0)$ & -99.4 & -11.7 & -36.0 & 5.3 & -28.30 & 0.00(2) & 0.52 \\
$^6$He & $(0^+;1)$ & -109 & -13.6 & -56 & 6.4 & -27.64 & -1.63(14) & 2.8 \\
$^6$Li & $(1^+;0)$ & -129 & -13.5 & -50 & 6.3 & -31.25 & -0.74(11) & 3.2 \\
$^7$He & $({\frac{3}{2}}^-;\frac{3}{2})$ & -110 & -14.1 & -61 & 6.7 & -25.2 & -3.7(2) & 4.7 \\
$^7$Li & $({\frac{3}{2}}^-;\frac{1}{2})$ & -153 & -17.1 & -68 & 8.2 & -37.4 & -1.8(3) & 4.7 \\
$^8$He & $(0^+;2)$ & -121 & -15.8 & -74 & 7.5 & -25.8 & -5.6(6) & 6.1 \\
$^8$Li & $(2^+;1)$ & -157 & -22.2 & -104 & 11.0 & -38.3 & -3.0(6) & 8.6 \\
$^8$Be & $(0^+;0)$ & -224 & -28.1 & -72 & 13.3 & -54.7 & -1.8(6) & 6.6 \\
\hline
\end{tabular}
\end{table}

Table 1 also reveals another problem; the difference $\Delta E_{VMC}$ 
between GFMC and variational Monte Carlo (VMC) energies
is surprisingly large in the p-shell nuclei.  Earlier successes in variational 
calculations of the $A \leq 4$ s-shell nuclei led to hopes that the 
variational calculations of larger nuclei and uniform nucleon matter, 
using cluster expansions \cite{PWP 92}, and 
chain summation methods \cite{AP 97} 
may have less than 10 \% errors.  The present VMC for $^8$Be, which 
includes many three-body correlations omitted in variational calculations 
of larger nuclei and uniform matter, has a 12 \% error.  It appears that 
important aspects of the wavefunctions of p-shell nuclei are still not
understood.

All NN interaction models are obtained by fitting NN
scattering data in the center of mass frame.
The model $v_{ij}$ denotes the 
interaction between two nucleons in the frame in which their total momentum 
${\bf P}_{ij} = {\bf p}_i + {\bf p}_j$, is zero.  In the rest frame of all 
nuclei other than the deuteron, the ${\bf P}_{ij}$ of a pair of nucleons 
is non-zero.  One then has to use the correct NN interaction:
\begin{equation}
v({\bf P}_{ij}) = v_{ij} + {\delta}v({\bf P}_{ij}),
\label{eq:vofp}
\end{equation}
between particles with total momentum ${\bf P}_{ij}$.  The correction 
${\delta}v({\bf P}_{ij})$ is called the boost interaction \cite{FPF 95}; 
it is zero when ${\bf P}_{ij} = 0$.

It is useful to consider a familiar example.  The Breit interaction 
\cite {BS 57} between two particles of mass $m$ and charge $Q$ is,
ignoring spin dependent terms for brevity, given by:
\begin{equation}
 \frac{Q^2}{r_{ij}} \left( 1 - \frac{{\bf p}_i \cdot {\bf p}_j}{2m^2} - \frac{{\bf p}_i \cdot {\bf r}_{ij} {\bf p}_j \cdot {\bf r}_{ij}}{2 m^2 r^2_{ij}}\right),
\label{eq:vemt}
\end{equation}
which depends upon both ${\bf p}_i$ and ${\bf p}_j$.  We can express it in
our notation as a sum of $v_{ij}$ and $\delta v({\bf P}_{ij})$:
\begin{eqnarray}
v_{ij} & = & \frac{Q^2}{r_{ij}} \left( 1 + \frac{p^2_{ij}}{2m^2} + \frac{({\bf p}_{ij} \cdot {\bf r}_{ij})^2}{2 m^2 r^2_{ij}}\right),
\label{eq:vemtd} \\
{\delta}v({\bf P}_{ij}) & = & -\ \frac{Q^2}{r_{ij}} \left( \frac{P^2_{ij}}{8m^2} + \frac{({\bf P}_{ij} \cdot {\bf r}_{ij})^2}{8 m^2 r^2_{ij}}\right),
\label{eq:vemb}
\end{eqnarray}
where ${\bf p}_{ij} = ({\bf p}_i - {\bf p}_j)/2$, is the relative momentum. 
The dependence of $v_{ij}$ on ${\bf p}_{ij}$ is included in all modern models, 
however, the ${\delta}v({\bf P}_{ij})$ has been 
neglected in the majority of nuclear and neutron star calculations.

Following the work of Krajcik and Foldy \cite{KF 74}, Friar \cite{F 75} 
obtained the following equation relating the boost interaction of order
$P^2$ to the interaction in the center of mass frame:
\begin{equation}
{\delta}v({\bf P}) = -\frac{P^2}{8m^2} v +\frac{1}{8m^2}[{\bf P \cdot r P \cdot {\nabla}}, v] +\frac{1}{8m^2}[({\sigma}_i - {\sigma}_j) \times {\bf P \cdot \nabla}, v] .
\label{eq:friar}
\end{equation}
The general validity of this equation in relativistic mechanics and field 
theory was recently discussed \cite{FPF 95}. 
  Nonrelativistic Hamiltonians containing boost interactions
include all terms quadratic in the particle velocities.
The contribution of the two-body boost interaction to the energy 
of light nuclei has been evaluated with VMC 
method \cite{CPS 93,FPCS 95,FPA 98}; 
it is repulsive and equals $\sim$ 37 \% of that of $V^R_{ijk}$ listed in 
Table 1.  Therefore 
about 37 \% of the $V^R_{ijk}$ in UIX 
simulates the contribution of the neglected ${\delta}v$. The three-nucleon 
interaction to be used in Hamiltonians containing ${\delta}v$ is denoted 
by $V^*_{ijk}$; the strength of $V^{*R}_{ijk}$ is 0.63 times that of 
$V^R_{ijk}$ in UIX, while $V^{*2\pi}_{ijk}=V^{2\pi}_{ijk}$.  In nuclei 
and in nuclear matter at densities up to $\sim 0.2$ nucleons/fm$^3$ 
Hamiltonians containing $V^*+\delta v$ and $V$ alone give rather similar 
results, but at higher densities they differ substantially.
Naturally the results of 
the Hamiltonian with $V^*+\delta v$ are more reliable.

One can also consider relativistic nuclear Hamiltonians of the type:
\begin{equation}
H_R = \sum \sqrt{p^2_i + m^2} + \sum ({\tilde v}_{ij} + {\delta}v({\bf P}_{ij})) + \sum {\tilde V}_{ijk} + ... ,
\label{eq:rhwb}
\end{equation}
where ${\tilde v}, {\tilde V}, ...$ include relativistic nonlocalities 
\cite{FPA 98}, and the boost interactions include terms of higher order in 
${\bf P}_{ij}$.  We must refit the two-nucleon scattering data 
to determine the ${\tilde v}$ in $H_R$, using relativistic kinetic energies 
\cite{CPS 93,FPA 98}.  Uncorrelated nucleons in nuclei have small momenta of 
order $m/4$, thus the correction to their nonrelativistic kinetic energy is 
negligible.  However, due to correlations induced by the strong ${\tilde v}$, 
a pair of nucleons acquire large relative momenta at small $r_{ij}$. 
The results obtained for light nuclei with $H_R$ and the $H$ given by 
eq.(\ref{eq:ham}) are very similar 
\cite{FPA 98}, indicating that substantial improvements are obtained by 
including the relativistic boost interactions in the nonrelativistic 
Hamiltonian, as is well known for electromagnetic interactions.

\section{Dense Nucleon Matter}

The equation of state (EOS) of cold symmetric nuclear matter (SNM)
and pure neutron matter (PNM) has been 
recently calculated with the variational method using chain summation 
techniques \cite{AP 97,APR 98}.  Calculations have been done with the 
Hamiltonians with and without boost interactions. 

\begin{figure}[htpb]
\centerline{\mbox{\epsfysize=7cm \epsffile{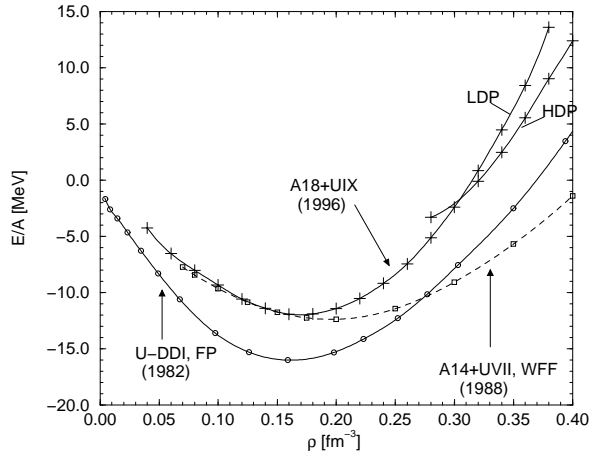}}}
\vspace{-0.5cm}
\caption{$E(\rho )$ of SNM calculated from A18 and UIX interactions compared with the results of earlier calculations.  The two sets of variational minima obtained at $\rho > 0.28/fm^3$ are labeled LDP and HDP for low and high density phases.}
\end{figure}

We will first discuss the results presented in ref. \cite{AP 97} for the
Hamiltonian without boost interactions.  The calculated SNM and PNM 
energies are shown in figs.3 and 4 
along with the results of earlier calculations \cite{FP 82,WFF 88}. 
The density dependence of the U-DDI interaction was choosen to obtain the 
empirical saturation properties of SNM, so the minimum of that curve may 
be regarded as experimental data.  The calculated energy of $-$12 MeV/A is 
higher than the observed $-$16 MeV/A.  However we now believe that much of 
this difference is due to the simplicity of the variational wave functions 
used in nucleon matter calculations.  Use of improved variational wave 
functions should lower the SNM energy by more that two MeV/A.  Comparison 
of VMC and GFMC results for energies of eight neutrons bound in a weak 
potential well \cite{PSCPPR 96} indicate that the present variational 
wave functions are more accurate for PNM than for SNM.  

\begin{figure}[htpb]
\centerline{\mbox{\epsfysize=7cm \epsffile{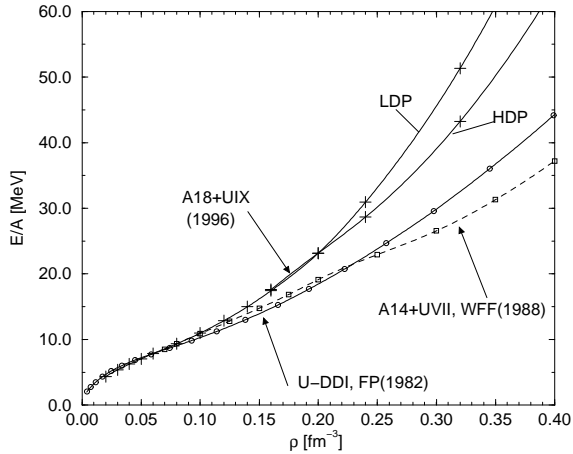}}}
\vspace{-0.5cm}
\caption{$E(\rho )$ of PNM. See fig.3 caption for details.}
\end{figure}

The calculated energies indicate a phase transition in SNM at a density 
of $\sim 0.3/fm^3$ (fig.3) and in PNM at $\sim 0.2/fm^3$ (fig.4).  Detailed 
analysis of the pair distribution functions and sums of response 
functions \cite{AP 97} indicate that the phase at higher densities has 
spin-isospin order expected from neutral pion condensation, first 
considered by Migdal \cite{MRMP}.  The 
variational wave functions used in this study are appropriate for 
isotropic matter.  The phase transition is signaled by a change 
in the range and strength of tensor correlations.  It is likely that 
the phase at higher density will be better described with a liquid 
crystal wave function with alternating spin layers \cite{PTPS} used in 
many studies with effective interaction models.  

As realized long ago by Migdal, this transition is very sensitive to the 
short range behavior of the ${\sigma}_i \cdot {\sigma}_j$ and 
${\sigma}_i \cdot {\sigma}_j {\tau}_i \cdot {\tau}_j$ interactions between 
nucleons.  It did not occur in either SNM or PNM with the Urbana $v_{14}$ 
model of 1981, while it occured only in PNM with the later Argonne $v_{14}$ 
model of 1984.  The Urbana-Argonne potentials have similar forms, but are fit 
to different data sets.  The 1981 (1984) Urbana (Argonne) $v_{14}$ models 
were fit to the n-p phase shifts available in the late 1970's (early 1980's),
while the A18 is fit directly to the Nijmegen 1994 
p-p and n-p scattering data base.  The quality of the fit obtained by the 
A18 potential is much higher, and thus it is likely that it 
provides a more accurate representation of the nuclear force.
The pion exchange part, 
$V^{2{\pi}}_{ijk}$, of the three-nucleon interaction is essential for the 
transition to occur in SNM, while in PNM it occurs at a much higher 
density in the absence of $V_{ijk}$.

The composition of nucleon matter energy calculated from A18 
and UIX interactions is listed at selected densities  
in table 2, in which T-1B denotes the Fermi gas kinetic energy.  The v-2B-S 
and T-2B-S list the two-body (2B) cluster contributions to the potential 
and kinetic energy from the static parts of the pair interaction
and correlation operators. 
The v-2B-MD and T-2B-MD show 2B contributions having one or more 
momentum dependent (MD) 
interaction or correlations.  These are conveniently regarded as the 
difference between the total 2B energies and their static parts.  The 
static many-body (MB) contributions, MB-S, as well as the expectation 
values of the three-nucleon interactions are calculated with the Fermi 
hypernetted and single operator chain summation methods, presumably quite 
accurately. However, the MB contributions containing MD interaction or 
correlations, as well as the boost interaction ${\delta}v({\bf P}_{ij})$, 
are evaluated from dressed three-body clusters.  The
${\delta}$E-2B lists an estimate of the change in energy that may be 
obtained by allowing additional flexibility in the two-body correlations.
We note that at high densities the MD parts, the $V^*_{ijk}$ and 
${\delta}v({\bf P}_{ij})$ give significant contributions.  

\begin{table}
\caption{Composition of nucleon matter energy in MeV.}
\vspace{0.2cm}
\begin{tabular}{lrrrrrrrr}
\hline
Type & SNM & SNM & SNM & SNM & PNM & PNM & PNM & PNM \\
$\rho (fm^{-3})$ & 0.08 & 0.16 & 0.32 & 0.64 & 0.08 & 0.16 & 0.32 & 0.64 \\
T-1B & 13.9 & 22.1 & 35.1 & 55.7 & 22.1 & 35.1 & 55.7 & 88.4 \\
v-2B-S & -36.5 & -66.7 & -117.9 & -227.2 & -26.7 & -49.4 & -93.0 & -185.7 \\
T-2B-S & 10.0 & 20.3 & 37.5 & 74.1 & 7.4 & 13.1 & 24.9 & 44.8 \\
v-2B-MD & 0.4 & 2.1 & 9.4 & 28.6 & 1.3 & 3.9 & 18.0 & 53.3 \\
T-2B-MD & 0.2 & 0.6 & 1.9 & 5.4 & 1.1 & 2.8 & 10.9 & 35.6 \\
MB-S & 3.4 & 5.5 & 9.2 & 25.6 & 3.2 & 6.4 & 1.8 & -9.0 \\
MB-MD & 0.7 & 3.2 & 13.4 & 61.7 & 0.1 & 1.1 & 14.2 & 60.1 \\
$V^{2{\pi}}_{ijk}$ & -0.8 & -3.6 & -13.3 & -82.0 & 0.3 & 1.2 & -17.4 & -76.9 \\
$V^{*R}_{ijk}$ & 0.9 & 4.0 & 19.4 & 98.2 & 0.5 & 2.8 & 19.3 & 101.0 \\
${\delta}v({\bf P}_{ij})$ & 0.6 & 2.1 & 6.4 & 21.0 & 0.7 & 2.2 & 6.9 & 21.5 \\
${\delta}$E-2B & -0.6 & -1.8 & -5.2 & -9.2 & -0.8 & -1.3 & -2.5 & -5.5 \\
Total E & -8.0 & -12.2 & -4.2 & 58.4 & 9.7 & 17.9 & 38.8 & 127.6 \\
\hline
\end{tabular}
\end{table}

\section{Neutron Star Matter}

The matter at subnuclear densities in the outer and inner crusts of neutron 
stars has interesting structures as discussed at this meeting 
by Pethick and Ravenhall 
\cite{PRHP,PR 95}.  Here we focus on the matter below the crust assuming 
that it is a uniform mixture of neutrons, protons, electrons and muons in 
beta equilibrium.  The calculated energies of SNM and PNM are fitted by 
generalized Skyrme type effective interactions \cite{PR 89} having different
parameters below and above the pion condensation phase transition.  The beta 
equilibrium conditions are calculated from these effective interactions.

Ignoring mixed phase regions the normal matter at density and proton fraction
of 0.204 fm$^{-3}$ and 0.073 is found to be in equilibrium with pion condensed 
matter at 0.237 fm$^{-3}$ and 0.057 for our most reliable model with boost 
and three-nucleon interactions.  Obviously the matter in the phase with
pion condensation has a lower charge density than the matter without
condensation.  Therefore, in reality the transition 
will proceed through mixed phase regions of the type discussed by Glendenning 
\cite{G 92} and Heiselberg {\em et. al.} \cite{HPS 93} in the context of the 
transition from hadronic to quark matter.   The mixed phase regions in the
pion condensation transition do not seem to have a large effect on the 
structure of neutron stars because the discontinuity in the charge density 
is rather small.  For example, the predicted thickness of the mixed phase 
regions varies from
$\sim$ 40 to 14 m in stars with 1.41 to 2.1 $M_{\odot}$; however 
these predictions are quite crude because our calculations of the pion 
condensed phase are still incomplete.

The calculated density dependence of the proton fraction in neutron star 
matter is shown in fig.5 for the Hamiltonians containing only A18, A18+
$\delta v$, A18+$\delta v$+UIX$^*$, and A18+UIX interactions.  The earlier 
results obtained with the U-DDI interaction (FPS) are 
shown for comparison.  Both the boost and the three 
nucleon interactions increase the proton fraction in matter,
but it remains below the critical value of 0.148 needed for direct Urca 
cooling, in the range of neutron star densities.  The plus signs 
in fig.5 show the proton fractions calculated with the A18 interaction 
alone with the LOB method \cite{EHMMP 97}.  These are in 
reasonable agreement with the results of our calculations up to a density 
of 0.6 fm$^{-3}$.  The density dependence of the electron chemical 
potential in matter is shown in fig.6. 

\begin{figure}[htpb]
\centerline{\mbox{\epsfysize=7cm \epsffile{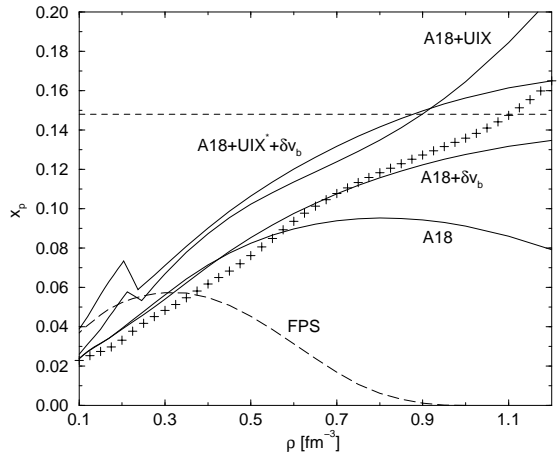}}}
\vspace{-0.5cm}
\caption{The proton fraction in matter in beta equilibrium for 
various model Hamiltonians.  The plus signs show the results obtained for 
the A18 Hamiltonian with LOB method.}
\end{figure}

\begin{figure}[htpb]
\centerline{\mbox{\epsfysize=7cm \epsffile{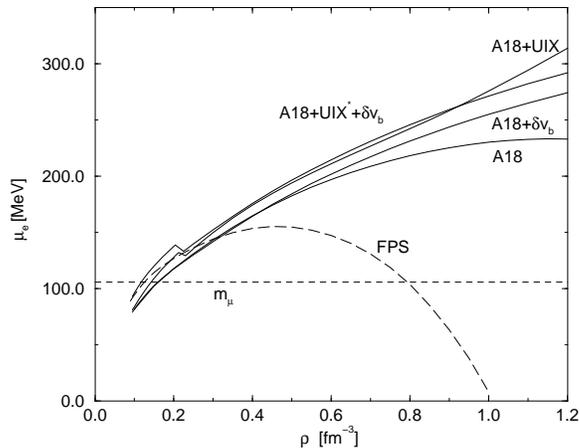}}}
\vspace{-0.5cm}
\caption{The electron chemical potential in neutron star matter for 
various model Hamiltonians.  Matter contains muons when $\mu_e$ 
exceeds the muon mass $m_{\mu}$.}
\end{figure}

The predicted neutron star properties appear in figures 7 and 8.  The 
dotted lines in these figures show results obtained using the PNM EOS.
They are not very different
from those obtained with matter in beta equilibrium. 
Due to the presence of momentum dependent and three-nucleon interactions in 
the present nuclear Hamiltonians, the predicted sound velocity in matter 
can exceed the velocity of light.  The densities at which superluminal 
sound occurs in each model are marked by vertical bars in fig.7.  These 
densities are very close to the maximum densities that can occur in neutron 
stars, therefore limiting the the velocity of sound to be $\leq c$ does 
not have a significant effect on the predicted maximum masses.

Observations of binary 
neutron stars have confirmed the existence of 1.4 $M_{\odot}$ neutron stars 
allowed by all models.  Recently several authors \cite{NSTA,NSTB,NSTC} have 
argued that there are indications of the existence of neutron stars 
with $M \sim 2 M_{\odot}$, which would rule out models without three-body 
forces, but see ref. \cite{NSTD} and Lamb's contribution \cite{LHIR}.

\begin{figure}[htpb]
\centerline{\mbox{\epsfysize=7cm \epsffile{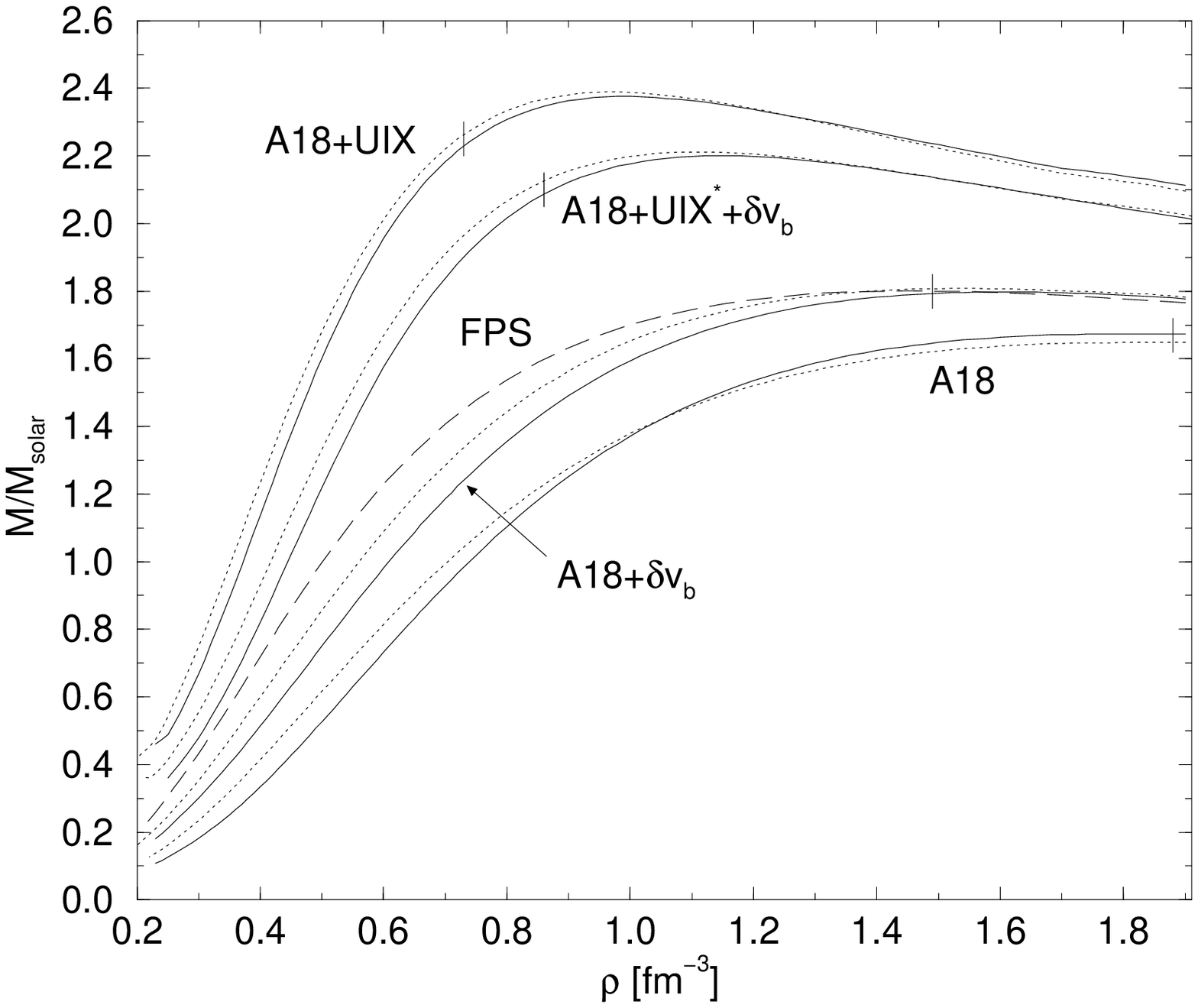}}}
\vspace{-0.5cm}
\caption{The dependence of neutron star mass on its central density for 
various model Hamiltonians. The full (dotted) lines show results obtained 
with the EOS of matter in beta equilibrium (PNM), and the 
vertical bars show where matter becomes superluminal in these models.}
\end{figure}

\begin{figure}[htpb]
\centerline{\mbox{\epsfysize=7cm \epsffile{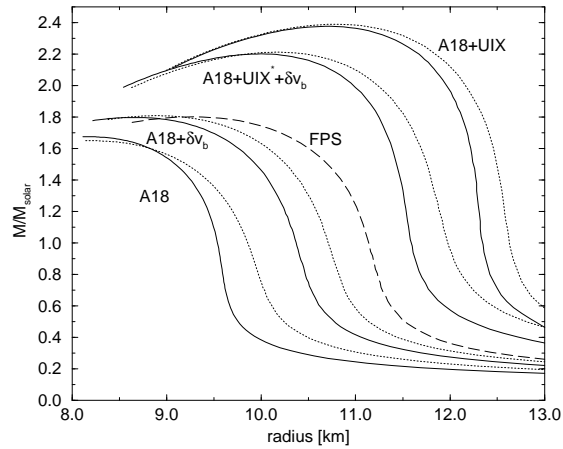}}}
\vspace{-0.5cm}
\caption{Neutron star mass-radius relation obtained from various model
Hamiltonians.  See caption of fig.7 for notation.}
\end{figure}

\section{Transition to Quark Matter}

It is expected that at some 
large density there will be a transition from nucleon matter to quark 
matter (QM).  Several authors \cite{FM 78} 
have studied the energy of cold QM
using the Bag-model, in which the total energy density 
contains a ``Bag-constant'' B, that takes into account the difference 
between the energies of the physical and QM vacua, and the energy of quarks 
interacting via one gluon exchange interaction calculated in first 
order of ${\alpha}_s$. The u and d quarks are assumed to be massless, 
and mass of s quarks is taken as 150 MeV.  The energy density of QM 
obtained with B = 122 and 200 MeV/fm$^3$ is compared with that of 
nucleon matter (NM) in beta equilibrium for the more realistic models 
containing boost interaction, in fig. 9. 
The value B = 122 MeV/fm$^3$ is supported by an analysis using the 
average of nucleon and delta-resonance masses \cite{CGS 86}.  
In the absence of three-body interactions NM is found to have lower energy 
than QM up to a large density of $\sim$ 1.6 fm$^{-3}$.  
However, the transition density is lowered to 
$\sim$ 1 fm$^{-3}$ after including the 
contributions of the three nucleon interaction.  

\begin{figure}[htpb]
\centerline{\mbox{\epsfysize=7cm \epsffile{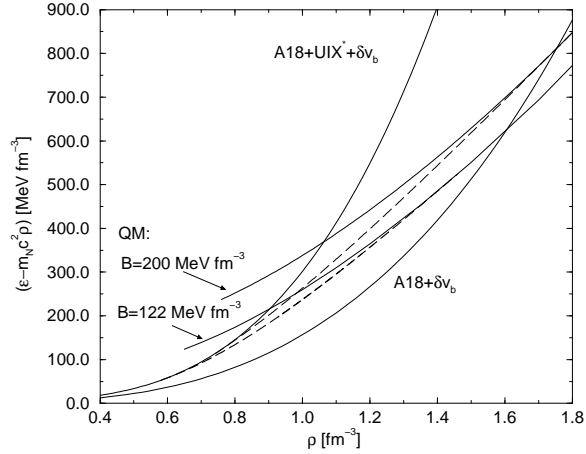}}}
\vspace{-0.5cm}
\caption{The energy densites of electrically neutral QM and NM are shown 
by full lines, while the dashed lines show those for matter with mixed 
QM and NM phases.}
\end{figure}

\begin{figure}[htpb]
\centerline{\mbox{\epsfysize=7cm \epsffile{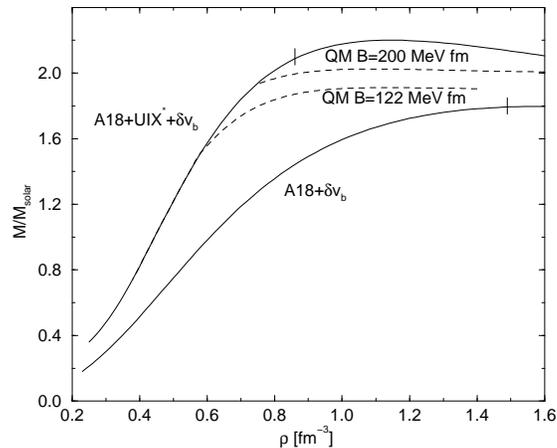}}}
\vspace{-0.5cm}
\caption{Dependence of neutron star mass on its central density.  Full lines 
show results obtained with NM EOS, while dashed lines show results 
obtained after including the effect of the transition to QM on the EOS.}
\end{figure}

Glendenning \cite{G 92} realized that it is not necessary to require 
the QM and NM phases to be separately charge neutral; 
a mixture of NM and QM can exist in a uniform lepton gas.
Neglecting the energy of the interface between the QM and NM, and using the 
A18 + $\delta v$ + UIX$^*$ model of NM, the transition occurs over the 
density range $\rho$ = 0.74 to 1.80 fm$^{-3}$ for B = 200, and 
0.58 to 1.46 fm$^{-3}$ for B = 122 MeV/fm$^3$.  At the lower end of this 
density range we have mostly NM with drops of QM, while at the higher 
end we will have mostly QM with drops of NM \cite{HPS 93}.

The dependence of the mass of neutron stars on their central density, 
obtained after incuding the effects of the transition to quark matter 
on the EOS are shown in fig.10.  For B = 122 Mev/fm$^3$ 
and the A18 + $\delta v$ + UIX$^*$ NM Hamiltonian stars with masses 
above 1.5 $M_{\odot}$ seem to have drops of quark matter in their cores.
The maximum mass is reduced to $\sim 1.9 M_{\odot}$ corresponding to a 
central density of $\sim 1.1$~fm$^{-3}$.  In this model pure quark matter 
appears at $\rho = 1.46$~fm$^{-3}$; therefore even the most massive stars 
have mixed QM and NM phases in their interior.  On the other hand, if 
B = 200 Mev/fm$^3$ the maximum mass becomes $\sim 2 M_{\odot}$, and 
stars with masses from 1.93 to 2.01 $M_{\odot}$ have quark drops in 
their interior.

\section*{Acknowledgments}
We would like to thank A. Arriaga, G. Baym, 
J. Carlson, J. Forest, H. Heiselberg, C. J. Pethick, 
S. C. Pieper, R. Schiavilla, E. F. Staubo and
R. B. Wiringa for many of the results discussed here, and M. Hjorth-Jensen
for communicating the results of neutron matter calculations.  This work has 
been partly supported by the US National Science Foundation under Grant 
PHY94-21309.


\eject

\end{document}